\documentstyle[twocolumn,prl,aps,epsfig]{revtex}
\newcommand{\be}{\begin{equation}}
\newcommand{\ee}{\end{equation}}
\newcommand{\bear}{\begin{eqnarray}}
\newcommand{\eear}{\end{eqnarray}}
\newcommand{\al}{\alpha}

\begin{document}
\draft
\wideabs{
\title{Exact soliton solutions, shape changing collisions and
partially coherent solitons in coupled
nonlinear Schr\"odinger equations}
\author{T. Kanna and M. Lakshmanan\cite{lakshman}}
\address{Centre for Nonlinear Dynamics, Department of Physics, Bharathidasan
University, Tiruchirapalli 620 024, India}
\date{}
\maketitle

\begin{abstract}
We present the exact bright one-soliton and two-soliton 
solutions of the integrable three coupled nonlinear Schr{\"o}dinger 
equations ($3$-CNLS) by using the Hirota method, and then
obtain them for the general $N$-coupled nonlinear Schr{\"o}dinger 
equations ($N$-CNLS). It is pointed out that the underlying solitons undergo
inelastic (shape changing) collisions due to intensity redistribution
among the modes. We also analyse the various possibilities and conditions for 
such collisions to occur. Further, we report the significant fact that
the various partial coherent solitons (PCS) discussed in the literature
are special cases of
the  higher order bright soliton solutions of the
$N$-CNLS equations.
\end{abstract}
}
In recent years the concept of soliton has been receiving
considerable attention in optical communications since soliton is 
capable of propagating over long distances without
change of  shape and with negligible attenuation [1-3].
It has been found that soliton propagation through
optical fiber arrays is governed by a set of equations 
related to the CNLS equations[1,2],
\be
iq_{jz}+q_{jtt}+2\mu\sum_{p=1}^N|q_p|^2q_j=0,\;\;j=1,2,...N,
\ee
where $q_j$ is the envelope in the $j$th core, $z$ and $t$ 
represent the
normalized distance along the fiber and the retarded time,
respectively.  Here $2\mu$ gives the strength of the nonlinearity.  Eq.
(1) reduces to the standard envelope soliton possessing integrable
nonlinear Schr{\"o}dinger equation  for $N=1$.  For $N=2$, the above
Eq. (1) governs the integrable Manakov system [4] and  recently for
this case the exact two-soliton solution has been obtained and novel
shape changing inelastic collision  property has been brought out [5].
However, the results are scarce for $N\geq 3$, even though the
underlying systems are of considerable physical interest. For example,
in addition to optical communication, in the context of bio-physics the
case $N=3$ can be used to study the launching and propagation of
solitons along the three spines of an alpha-helix in protein[6].
Similarly the CNLS Eq. (1) and its generalizations for $N$$\geq$$3$ are
of physical relevance in the theory of soliton wavelength division
multiplexing[7], multi-channel bit-parallel-wavelength optical fiber
networks[8] and so on. In particular, for arbitrary $N$, Eq. (1) governs
the propagation of $N$-self trapped mutually incoherent wavepackets in
Kerr-like photorefractive media[9] in which $q_j$ is the $j$th
component of the beam, $z$ and $t$ represents the normalized 
coordinates along the
direction of propagation and the transverse coordinate, respectively, and 
$\sum_{p=1}^N|q_p|^2$ represents the change in the refractive index profile 
created by all incoherent components of the light beam [9] when 
the medium response is slow.The parameter $\mu$ $=$ 
$k_0^2 n_2/2$, where $ n_2$ is the nonlinear Kerr coefficient and 
$k_0$ is the free space wave vector. 

In this letter, we report the exact bright one and two
soliton solutions, first for the $N=3$ case and then
for the arbitrary $N$ case,
where the procedure can be extended in principle to higher order soliton solutions,
using the Hirota bilinearization method. In particular, we
point out that the shape changing inelastic collision property
persists for the $N$ $\geq$ $3$ cases also as in the 
$N=2$ (Manakov) case reported recently [5], 
giving rise to many possibilities of energy exchange.
Furthermore, we point
out that in the context of spatial solitons the partially coherent
stationary solitons(PCS) reported in
the recent literature[9-10] are  special cases of the above general
soliton solutions which  undergo shape changing collisions. 

The bright one-soliton and two-soliton solutions of the $3$-CNLS system,
\bear
iq_{jz}+q_{jtt}+2\mu(|q_1|^2+|q_2|^2+|q_3|^2)q_j=0,\;j=1,2,3,
\eear
can be obtained from its equivalent bilinear form 
resulting from the transformation $q_j$ $=$ $g^{(j)}/f$,
\bear
(iD_z+D_t^2)g^{(j)}.f= 0,\;\;
D_t^2f.f= 2\mu\sum_{n=1}^3{g^{(n)}g^{(n)*}},
\eear
where $*$ denotes the complex conjugate, $D_z$ and $D_t$ are Hirota's
bilinear operators [11], and $g^{(j)}$'s  are complex functions, while
$f(z,t)$ is a real function.The resulting set of Eqs. (3) can be solved
recursively by making the power series expansion
$g^{(j)}=\chi g_1^{(j)}+ \chi^3 g_3^{(j)}+\ldots$,
$f=1+\chi^2 f_2+\chi^4 f_4+\ldots$,
$j=1,2,3$, where $\chi $ is a formal expansion parameter.  In order to
get the one-soliton solution, the power series expansions are
terminated as
$g^{(j)}=\chi g_1^{(j)}$, and $f=1+\chi^2 f_2$.
After deducing $g^{(j)}$ and $f$ as 
$ g^{(j)} = \al_1^{(j)} e^{\eta_1}$, $j=1,2,3$ and 
$f=1+e^{\eta_1+\eta_1^*+R}$, 
where $e^R=\mu\sum_{j=1}^3{|\alpha_1^{(j)}|^2}/(k_1+k_1^*)^2,$
the bright one-soliton solution is obtained as
\bear
\left(
q_1,
q_2, 
q_3
\right)^T
& = &
\frac{e^{\eta_1}}{1+e^{\eta_1+\eta_1^*+R}}\left(
\al_1^{(1)},
\al_1^{(2)},
\al_1^{(3)}
\right)^T,\nonumber\\
& = &
\frac{k_{1R}e^{i\eta_{1I}}}{\mbox{cosh}
\left(\eta_{1R}+\frac{R}{2}\right)}\left(
A_1,
A_2,
A_3
\right)^T,
\eear
where $\eta_1$$=$$k_1(t+ik_1z)$,$\;$$A_j$$=$$\al_1^{(j)}/
\Delta$
 and
$\Delta=
(\mu(\sum_{j=1}^3{|\al_1^{(j)}|^2}))^{1/2}$. 
Here $\al_1^{(j)}$, $k_1$, $j=1,2,3$, are four arbitrary complex
parameters. Further $k_{1R}A_j$ gives the amplitude of the $j$th mode
 and $2k_{1I}$ the soliton velocity.

The general bright two-soliton solution of Eq. (2) can be generated
by terminating the series as $g^{(j)}=\chi g_1^{(j)} + \chi^3 g_3^{(j)}$
and $f=1+\chi^2 f_2 +\chi^4 f_4$ and solving the resultant
linear partial differential equations.
This solution contains eight arbitrary complex parameters,
$\al_l^{(j)}$ and  $k_l$, $l=1,2$ and $j=1,2,3$. It is given by
\bear
q_j&=&(\al_1^{(j)}e^{\eta_1}+\al_2^{(j)}e^{\eta_2}+e^{\eta_1+\eta_1^*+\eta_2+\delta_{1j}}\nonumber\\
&&+e^{\eta_1+\eta_2+\eta_2^*+\delta_{2j}})/D,\;j=1,2,3,
\eear
where $D=1$ $+$ $e^{\eta_1+\eta_1^*+R_1}$ $+$ $e^{\eta_1+\eta_2^*+\delta_0}$
$+$ $e^{\eta_1^*+\eta_2+\delta_0^*}$ $+$ $e^{\eta_2+\eta_2^*+R_2}$
$+$ $e^{\eta_1+\eta_1^*+\eta_2+\eta_2^*+R_3}$, 
$\eta_i$$=$$k_i(t+ik_iz)$,
$e^{R_1}=$ $\kappa_{11}/(k_1+k_1^*)$,
$e^{R_2}=$ $\kappa_{22}/(k_2+k_2^*)$,
$e^{\delta_0}=$ $\kappa_{12}/
(k_1+k_2^*)$,
$e^{\delta_{1j}}=$ $((k_1-k_2)(\al_1^{(j)}\kappa_{21}-\al_2^{(j)}
\kappa_{11}))/
((k_1+k_1^*)(k_1^*+k_2))$,
$e^{\delta_{2j}}=$ $((k_2-k_1)(\al_2^{(j)}\kappa_{12}-\al_1^{(j)}\kappa_{22})
)/((k_2+k_2^*)(k_1+k_2^*))$,
$e^{R_3}=$ $(|k_1-k_2|^2(\kappa_{11}\kappa_{22}-\kappa_{12}\kappa_{21})
)/((k_1+k_1^*)(k_2+k_2^*)|k_1+k_2^*|^2)$
and
$\kappa_{il}=$ $\mu\sum_{n=1}^{3}{\alpha_i^{(n)}\alpha_l^{(n)*}}
/(k_i+k_l^*)$, $i,\;l=1,2$, $j=1,2,3$.
Though one  can proceed to obtain higher order soliton solutions in
principle by making use of the general power series expansion, the
details become complicated and we will present the results
separately.

The nature of the  interaction of the underlying solitons
can be well understood by making an
asymptotic analysis of the two-soliton solution[5].
Asymptotically, the two-soliton solution(5) can be written as
a combination of two one-soliton solutions and their forms in the two different 
regimes $z \rightarrow -\infty$ and $z \rightarrow\infty$ are similar to those
of the one-soliton solution given in Eq. (4) but differing in
amplitude (intensity)
and phase. The analysis reveals that there is an intensity exchange 
among the three components of each soliton during this 
two-soliton interaction, which can be quantified by defining a transition matrix
$T_j^l$ such that
$A_j^{l+}$$=$$A_j^{l-}T_j^l$,$\;\;$$j$$=$$1,2,3$ and $l$$=$$1,2,$
where the superscripts $l\pm$ represent 
the solitons designated as $S_1$ and $S_2$ at $z\rightarrow \pm\infty$
, and $k_{lR}A_j^{l\pm}$denote the  corresponding 
amplitudes.

Consequently, the three modes $q_1$, $q_2$ and $q_3$ of $S_1$
having magnitudes of amplitudes
$|A_j^{1-}|k_{1R}=$$|\al_1^{(j)}|k_{1R}/\Delta_1$,
where
$\Delta_1$$=$$(\mu(\sum_{j=1}^3{|\al_1^{(j)}|^2}))^{1/2}$,
exchange intensity given by the square of the transition matrices,
$|T_j^{1}|^2=|1-\lambda_2(\al_2^{(j)}/\al_1^{(j)})|^2/
|1-\lambda_1\lambda_2|,$$\;$$j=1,2,3,$
 along with a phase shift  
$\Phi^1=(R_3-R_2-R_1)/2$ during
collision. Here $\lambda_1=\kappa_{21}/\kappa_{11}$
and $\lambda_2=\kappa_{12}/\kappa_{22}$.
In a similar fashion due to collision the three modes $q_1$, $q_2$ and $q_3$
of $S_2$ also exchange an amount of intensity,
$|T_j^{2}|^2=|1-\lambda_1\lambda_2|/
|1-\lambda_1(\al_1^{(j)}/\al_2^{(j)})|^2
$,$\;$$j=1,2,3,$
respectively and change their amplitudes to
$|A_j^{2+}|k_{2R}=|\al_2^{(j)}|k_{2R}/\Delta_2$
from
$|A_j^{2-}|k_{2R},$ 
respectively. Here $\Delta_2$$=$$(\mu(\sum_{j=1}^3
{|\al_2^{(j)}|^2}))^{1/2}$.
The associated phase shift for this soliton is 
$\Phi^2=-(R_3-R_2-R_1)/2.$ We also note there is a 
net change in the  
relative separation distance between the solitons due to collision by 
$\Delta x_{12}=(k_{1R}+k_{2R})|(R_3-R_2-R_1)|/2k_{1R}k_{2R}.$

Also, we note that for the special case $|T_j^l|=1$, $l=1,2$,
$j=1,2,3$, which is possible only when
$\al_1^{(1)}/\al_2^{(1)}$= $\al_1^{(2)}/\al_2^{(2)}$= 
$\al_1^{(3)}/\al_2^{(3)}$, 
the collision corresponds to the standard elastic collision.  For all
other cases, the quantity $|T_j^l|\neq 1$, which corresponds to a
change in the values of the amplitudes of the individual modes leading
to a redistribution of the intensities among them and corresponding to
a change in the shape of the soliton.  However, during the interaction
the total intensity of the individual solitons $S_1$ and $S_2$ remains
conserved, that is
$|A_1^{l\mp}|^2+|A_2^{l\mp}|^2+|A_3^{l\mp}|^2=1/\mu$.

The above shape changing (inelastic) collision during the two-soliton
interaction of the $3$-CNLS can occur in two different ways. The first
case is an enhancement of intensity in anyone of the modes of either
one of the solitons (say $S_1$) and suppression in the remaining two
modes of the corresponding soliton with commensurate changes in the
other soliton $S_2$.  The other possibility is an interaction which
allows one of the modes of either one of the solitons (say $S_1$) to
get suppressed while the other two modes of the corresponding soliton
to get enhanced (with corresponding changes in $S_2$). In either of the
cases, the intensity may be completely or partially
suppressed(enhanced).  Thus as a whole during the inelastic interaction
among the two one solitons $S_1$ and $S_2$ of the $3$-CNLS, the soliton
$S_1$ ($S_2$)  has the following six possible combinations to exchange
the intensity among its modes:
$(q_1, q_2, q_3)\rightarrow(q_1^a, q_2^b, q_3^c)_i,
(a,b,c = S\,\text{(suppression)},E\,\text{(enhancement)})$ 
with  $i=1$, $a=E$, $b=S$, $c=S$; $i=2$, $a=S$, $b=E$, $c=S$; $i=3$,
$a=S$, $b=S$, $c=E$; $i=4$, $a=S$, $b=E$, $c=E$; $i=5$, $a=E$, $b=S$,
$c=E$ and $i=6$, $a=E$, $b=E$, $c=S$.

Two of the above interactions involving a dramatic switching 
in the intensity are depicted in Fig. 1 for illustrative
purpose for specific choice of soliton parameters. 
These may also be viewed as the two-soliton interaction 
in a waveguide supporting propagation of three nonlinear waves 
simultaneously. For other choices, in general, partial 
suppression(enhancement) of intensity among the components will 
occur depending on the values of the transition matrix elements 
$T_j^l$. Fig. 1a is plotted
for the parameters $k_1=1+i$, $\;$ $k_2=2-i$,
$\alpha_2^{(1)}=\alpha_2^{(2)}=(39+i80)/89$,
$\alpha_1^{(1)}=\alpha_1^{(2)}=\alpha_1^{(3)}=\alpha_2^{(3)}=1$ and $\mu=1$.
In this figure the intensities of the components $q_1$ and $q_2$ of
$S_1$ ($S_2$) are almost completely suppressed (enhanced) and that of the
third component is enhanced (suppressed). The second possibility of
enhancement (suppression) of intensity in the $q_1$ and $q_2$
components of $S_1$($S_2$) and suppression(enhancement) of intensity in
its $q_3$ component are shown in Fig. 1b, in which the parameters are
chosen as
$k_1=1+i$, $k_2=2-i$, 
$\alpha_1^{(1)}=0.02+0.1i$,
$\alpha_1^{(2)}=0.1i$,
$\alpha_1^{(3)}=\alpha_2^{(1)}=\alpha_2^{(2)}=\alpha_2^{(3)}=1$.

Now it is  straight forward  to extend the above analysis to obtain the
one-soliton and two-soliton solutions of the arbitrary $N$-CNLS Eq. (1).
After making the bilinear transformation $q_j$ $=$ $g^{(j)}/f$,
$j=1,2,3,\ldots,N$ in Eq. (1), one can get a set of bilinear equations of
the form (3) but now with $j,\; n=1,2,3,\ldots,N$.  Then by expanding
$g^{(j)}$s and $f$ in power series up to $N$terms and following the
procedure mentioned above, the one-soliton and two-soliton solutions of
Eq. (1) can be obtained.

\emph{(a) One-soliton solution:}
\begin{eqnarray}
\left(
q_1,
q_2,\ldots, 
q_N
\right)^T
& = &
\frac{k_{1R}e^{i\eta_{1I}}}{\mbox{cosh}\,(\eta_{1R}+\frac{R}{2})}\left(
A_1,
A_2,\ldots,
A_N
\right)^T
, 
\end{eqnarray}
where 
$\eta_1=k_1(t+ik_1z)$,
$A_j =\alpha_1^{(j)}/\Delta$,
$\Delta = (\mu(\sum_{j=1}^N{|\alpha_1^{(j)}|^2}))^{1/2}$,
$e^R=\Delta^2/(k_1+k_1^*)^2$,
$\alpha_1^{(j)}$ and $k_1$, 
$j$$=$$1,2\ldots,N,$ are $(N+1)$ arbitrary complex parameters.

\emph{(b) Two-soliton solution:}
This solution will also be of the same form as Eq.(5)
with the replacements, $j=1,2,\ldots,N$ and  
$\kappa_{il}$ $=$ $\mu\sum_{n=1}^N{
\alpha_i^{(n)}\alpha_l^{(n)*}}/(k_i+k_l^*)$,
where $i,\;l=1,2$. One can also verify that this
two-soliton solution will depend on $2(N+1)$ complex parameters and
the shape changing interaction can lead to intensity
redistribution among the modes of the soliton of the $N$-CNLS system in
$2^N-2$ ways (by generalizing the $N=3$ case).  We believe that such
studies will have important applications in logic gates and 
all optical computations[12].

From an application point of view, it has been observed recently that
the CNLS equations can support a kind of stationary solutions known as
partially coherent solitons (PCS). In particular, explicit form of such
solutions have been given for $N=2,3$ and $4$ cases of Eq. (1) in
Ref.[9]. They have also been shown to have variable shapes. Now, the
generalized Manakov equation(1) is integrable[13] and hence its
$N$-soliton solution can be obtained in principle by extending our above
analysis. So the natural question arises as to what is the relation
between the PCS and the exact $N$-soliton solutions.

To answer the above question, let us look at the $N=2,3$ and $4$, 
cases
of Eq. (1) explicitly. One can check that very special cases
corresponding to specific parametric restrictions in the two-soliton
solution of the $N=2$ case, the three-soliton solution of the $N=3$
case and four-soliton solution of the $N=4$ case give rise to the
$2$-soliton, $3$-soliton and the $4$-soliton PCSs, respectively,
reported in [9]. In order to appreciate this we consider as an
illustration the three-soliton solution of the $N=3$ case of Eq. (1).
Instead of writing down the full $3$-soliton solution of the $N=3$ case
explicitly and choosing the special parametric values, we make the
following simplified procedure.  Starting from the bilinear Eqs. (3)
and terminating the series for $g^{(j)}$ and $f$ as
$g^{(j)}=\chi g_1^{(j)}+\chi^3 g_3^{(j)}+\chi^5 g_5^{(j)}$ 
and $f= 1+\chi^2 f_2+\chi^4 f_4+\chi^6 f_6$,
one can identify 
$g_1^{(j)}=\alpha_1^{(j)}e^{\eta_1}+\alpha_2^{(j)}e^{\eta_2}
+\alpha_3^{(j)}e^{\eta_3}$
where $\eta_n=k_n(t+ik_nz)$, $j,\;n = 1,2,3$ in which $\alpha_i^{(j)}$
and $k_i$ are complex parameters.  Finding $g_3^{(j)}$, $g_5^{(j)}$,
$j=1,2,3$, $f_2$, $f_4$ and $f_6$, the three soliton solution is
obtained. Instead, as a special case, we look for a stationary solution
with $k_{nI}=0$,
$\alpha_2^{(1)}=\alpha_3^{(1)}=\alpha_1^{(2)}=\alpha_3^{(2)}=
\alpha_1^{(3)}=\alpha_2^{(3)}=0$ and 
$\alpha_1^{(1)}=-\alpha_2^{(2)}=\alpha_3^{(3)}=1$, 
in order to gain insight into the physics of the problem.  Then, the
resulting explicit expression for the three-soliton solution 
has been found 
after simple algebraic manipulation to be exactly the same as the 
stationary PCS for N=3 given in Eq.(17) of Ref.[9a]. One can also
check that with the choice $k_{nI}=0$, $\alpha_2^{(1)}=\alpha_1^{(2)}=0$,
$\alpha_1^{(1)}=-\alpha_2^{(2)}=1$, in the two-soliton solution of the
$N=2$ case (Manakov equation) [5] of Eq. (1), the $N=2$ soliton
complex(PCS) is obtained.  By a similar analysis we have verified that
the $N=4$ PCS also results as a special case of the $4$-soliton
solution of the $4$-CNLS equation.  Extending this idea it is clear
that the PCS which is formed due to a nonlinear superposition of
$N$-fundamental solitons[9] is a special case of $N$-soliton solution
of the $N$-CNLS Eq. (1).

Further, it has been found that these PCS are of variable shape and
also change their shape during collision with another PCS [9]. The
reason for the shape change of
the PCS can be deduced from the
interaction properties of the solitons discussed above.
The solitons are characterized by their amplitudes $A_j^lk_{lR}$ and their 
velocities $2k_{lI}$(so that the angle of incidence is 
$\theta_l=tan^{-1}(2k_{lI}))$. 
During a pair-wise interaction of two fundamental solitons of
$N$-CNLS equation there is an energy sharing between them 
resulting in a novel shape changing collision, depending 
on the transition matrix elements $T_j^l$, the phase shift $\Phi^l$ and 
change in the relative separation distance $\Delta x_{ij}$ defined earlier
. Since the PCS is a special case of
the $N$-soliton solution, parametrized as above,
it naturally possesses a variable shape. 
For example in Figs.1, the solitons $S_1$ and $S_2$ are
travelling with 
velocities $2k_{1I}=2$ and $2k_{2I}=-2$, respectively. For the chosen 
parameters, in Fig.1a, the amplitudes of the modes of the solitons
$S_1$ and $S_2$
before interaction given respectively by $k_{1R}$$|A_j^{1-}|$ $=$
$(0.577,0.577,0.577)$ and $k_{2R}|A_j^{2-}|$
$=$ $(0.857,0.857,1.591)$ 
change to 
$k_{1R}|A_j^{1+}|$ $=$
 $(0.183,0.183,0.966)$
and $k_{2R}|A_j^{2+}|$ $=$ $(1.155,1.155,1.155)$
after interaction,  preserving the total intensity of 
each of the soliton. Similarly in Fig.1b, the amplitudes of the solitons 
$S_1$ and $S_2$ 
before interaction are $(0.101,0.099,0.990)$ and $(1.322,1.335,0.686)$ 
respectively, 
while after interaction they become $(0.466,0.484,0.741)$ and 
$(1.155,1.155,1.155)$.The phase shifts suffered by the solitons are 
$\Phi^1$ $=$ $-\Phi^2$ $=$ $-0.787$(Fig.1a),$-0.600$(Fig.1b). 
During the collision 
process the initial separation distance  $x_{12}^-=-0.668$ 
changes to $x_{12}^+=0.513$ in Fig.1a and $-0.036$ to $0.865$ in Fig.1b.
\begin{figure}[h]
\centerline{\epsfig{figure=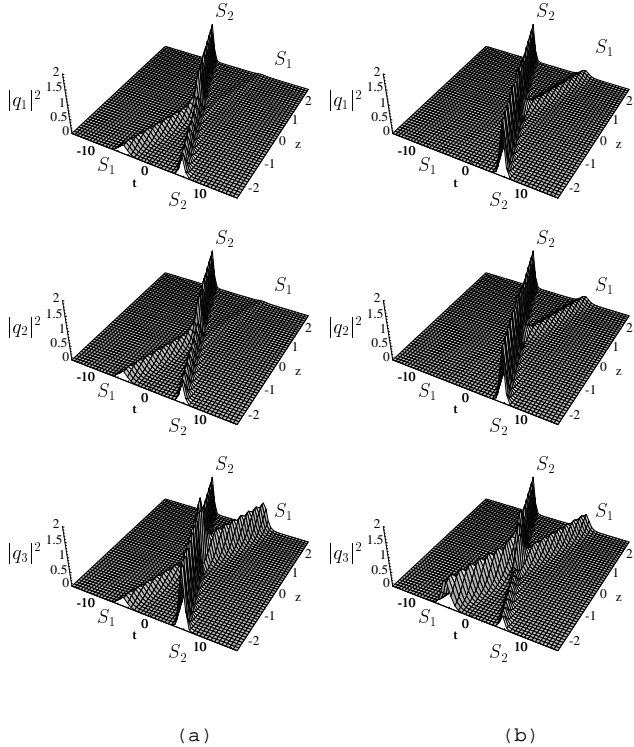, width=\columnwidth}}
\caption{Intensity profiles of the three modes of the two-soliton 
solution in a waveguide described by the  CNLS Eq. (2) showing 
two different  dramatic scenarios of the shape changing collision. 
The parameters 
are chosen as in the text.}
\end{figure}
In a similar way, the variable shape of the PCS during interaction with
another PCS also arises from the fundamental bright soliton collision
of the Manakov system.
The collision of two PCS each comprising of m
and n soliton complexes respectively, such that $m+n=N$, is equivalent
to the interaction of $N$ fundamental bright solitons (for suitable
choice of parameters) represented by the special case of
$N$-soliton solution of the $N$-CNLS system.Further details will be 
published elsewhere.

Our above analysis has  
considerable practical relevance in view of the various recently 
reported interesting experimental observations. Firstly, the 
Manakov spatial solitons have been observed in AlGaAs planar 
waveguides[14] and their collisions involving energy exchange of precisely 
the type discussed here has 
been experimentally demonstrated[15]. Also collisions between PCS's 
of shape changing type as treated here were 
observed in a photorefractive strontium barium niobate crystal 
using screening solitons[16]. Further partially incoherent solitons 
have been observed through excitation by partially coherent 
light[17] and with a light bulb[18]. Using different techniques, 
such as the coherent density function theory[19], to describe such 
incoherent solitons one can obtain the N-coupled NLS equations of 
the form(1) considered in this Letter. We believe that our exact 
analytical results will give further impetus in the experimental 
investigations of these solitons.

In conclusion, we have shown that  $N$-CNLS Eq. (1) possesses
fascinating type of soliton solutions undergoing shape changing
(inelastic) collision property due to intensity redistribution among
its modes. The many possibilities for such collisions to occur provides
interesting avenues of research in developing logic gates and in
all-optical digital computations[12]. We have also shown the
interesting fact that the multisoliton complexes are special cases of
the shape changing bright soliton solutions and pointed out that the
variable shape of PCS is due to the shape changing collision of the
fundamental solitons which is an inherent nature of the $N$-CNLS system.

The work of M.L. and T.K. forms part of a Dept. of Science
and Technology, Govt. of India research project.


\begin{thebibliography}{99}
\bibitem[*]{lakshman}
Electronic address: lakshman25@satyam.net.in

\bibitem{a}
G. P. Agrawal, {\it Nonlinear Fiber Optics-Second Edition} 
(Academic Press, New York, 1995).

\bibitem{b}
N. Akhmediev and A. Ankiewicz, {\it Solitons: Nonlinear Pulses and Beams} 
(Chapman and  Hall, London, 1997).

\bibitem{e}
A. Hasegawa and Y. Kodama, {\it Solitons in Optical Communications}
(Oxford University Press, England, 1995).

\bibitem{d}
S. V. Manakov, Zh. Eksp. Teor. Fiz. {\bf 65}, 505 (1973) [Sov. Phys.
JETP {\bf 38}, 248 (1974)].

\bibitem{i}
R. Radhakrishnan, M. Lakshmanan, and J. Hietarinta, Phys. Rev.  E
{\bf56}, 2213 (1997).

\bibitem{j}
A. C. Scott, Phys. Scr. {\bf 29}, 279 (1984).

\bibitem{j1}
S. Chakravarty, M. J. Ablowitz, J. R. Sauer, and R. B.
Jenkins, Opt. Lett. {\bf 20}, 136 (1995); 
\bibitem{j2}
C. Yeh and L. Bergman, Phys. Rev. E {\bf 57}, 2398 (1998).

\bibitem{p}
[a] A. Ankiewicz, W. Krolikowski, and N. N. Akhmediev, Phys. Rev. E 
{\bf 59}, 6079 (1999); 
[b] N. Akhmediev, W. Krolikowski, and A. W. Snyder, Phys. Rev. Lett.
{\bf 81}, 4632 (1998).

\bibitem{l}
V. Kutuzov, V. M. Petnikova, V. V. Shuvalov, and V. A. Vysloukh,
Phys. Rev. E {\bf 57}, 6056 (1998).

\bibitem{m}
R. Hirota, J. Math. Phys. (N.Y.) {\bf 14}, 805 (1973).

\bibitem{n}
M. H. Jakubowski, K. Steiglitz and R. Squier,
Phys. Rev. E {\bf 58}, 6752 (1998).

\bibitem{o}
V. G. Makhan'kov and O. K. Pashaev, Theor. Math. Phys. {\bf 53}, 979 (1982);
K. Nakkeeran, Phys. Rev. E {\bf 62}, 1313 (2000).


\bibitem{p}
J. U. Kang, G. I. Stegeman, J. S. Aitchison, and N. Akhmediev, 
Phys. Rev. Lett. {\bf 76}, 3699 (1996).
\bibitem{q}
C. Anastassiou {\it et al}., 
Phys. Rev. Lett. {\bf 83}, 2332 (1999).
\bibitem{r}
W. Krolikowski, N. Akhmediev, and B. Luther-Davies, 
Phys. Rev. E {\bf 59}, 4654 (1999).
\bibitem{r}
M. Mitchell, Z. Chen, M. F. Shih, and N. Segev, 
Phys. Rev. Lett.  {\bf 77}, 490 (1996).
\bibitem{s}
M. Mitchell and M. Segev, 
Nature, {\bf 387}, 880 (1997).
\bibitem{t}
D. N. Christodoulides et al.,
Phys. Rev. Lett.  {\bf 78}, 646 (1997);  D. N. Christodoulides, 
T. H. Coskun, and R. I. Joseph, Opt. Lett. {\bf 22}, 1080 (1997).



\end{thebibliography}
\end{document}